# Interplay between intra- and inter-nanowires dynamic dipolar interactions in the spin wave band structure of Py/Cu/Py nanowires

G. Gubbiotti,[1] X. Zhou,[2] Z. Haghshenasfard,[3] M. G. Cottam,[3] A.O. Adeyeye[2] and M. Kostylev[4]

[1]Istituto Officina dei Materiali del CNR (CNR-IOM), Sede Secondaria di Perugia, c/o Dipartimento di Fisica e Geologia, Università di Perugia, I-06123 Perugia, Italy

[2]Information Storage Materials Laboratory, Department of Electrical and Computer Engineering, National University of Singapore, 117576 Singapore.

[3] University of Western Ontario, Department of Physics and Astronomy, London, Ontario N6A 3K7, Canada

[4]School of Physics M013, University of Western Australia, 35 Stirling Hwy, 6009 Western Australia, Australia

ABSTRACT

We have studied both experimentally and theoretically the reprogrammable spin wave band structure in Permalloy(10nm)/Cu(5nm)/Permalloy(30nm) nanowire arrays of width *w*=280 nm and inter-wire separation in the range from 80 to 280 nm. We found that, depending on the inter-wire separation, the anti-parallel configuration, where the magnetizations of the two Permalloy layers point in opposite directions, is stabilized over specific magnetic field ranges thus enabling us to directly compare the band structure with that of the parallel alignment. We show that collective spin waves of the Bloch type propagate through the arrays with different magnonic bandwidths as a consequence of the interplay between the **intra**- and **inter**-nanowire dynamic dipolar interactions. A detailed understanding, e.g. whether they have a stationary or propagating character, is achieved by considering the phase relation (in-phase or out-of-phase) between the dynamic magnetizations in the two ferromagnetic layers and their average value. This work opens the path to magnetic field-controlled reconfigurable layered magnonic crystals that can be used for future nanoscale magnon spintronic devices.

## 1. INTRODUCTION

In the last few years, reprogrammable magnonic crystals (MCs) whose dynamic response can be changed on demand have been demonstrated in planar one- and two-dimensional patterned nanostructures composed of two sub-units which exhibit different reversal field and can thus be switched independently one from another. [1,2,3,4,5,6,7,8,9,10,11,12,13,14,15,16] In addition, promising functionalities have been demonstrated for dynamically controllable MCs, whose reconfiguration time is small in comparison with the relevant spin-wave processes. [17,18,19] To overcome the limitations of conventional lithography, which is very susceptible to defects and changes of the magnetic properties near the edges thus affecting the spin wave (SW) propagation, researchers have proposed alternative methods to create periodic magnetic nanostructures.[20] For example, the magnetization modulation in magnetic films without morphological patterning have been reconfigurably created by ion-implantation,[21] laser annealing [4] and thermally assisted scanning probe lithography.[5] In addition, reconfigurable magnonic conduits were realized using domain walls to channel and bend SW beams thus promising complex and reconfigurable wiring in spin-wave-based circuits and paving the way toward domain wall nanoelectronics.[22] In this respect, programmable SW filtering has also been experimentally and numerically demonstrated by resetting the spin structure of pinned 90° Néel domain walls in a continuous CoFeB film with abrupt rotations of uniaxial magnetic anisotropy.[23]

Another promising approach to reconfigurable magnonic band structure (MBS) is based on the exploitation of the vertical dynamic coupling by the stray dipolar field in layered magnonic crystals, i.e. a continuous film and an array of magnetic dots or wires deposited on top or by using layered arrays of magnetic elements with or without a non-magnetic spacer. [24,25,26,27,28,29,30] Following this idea, we have recently demonstrated the possibility of controlling the MBS by switching between different magnetization configurations, parallel (P) and anti-parallel (AP), in dense arrays of Py/Cu(10nm)/Py nanowires (NWs) with different separations and where the two Py (Permalloy, $Ni_{80}Fe_{20}$) layers have different thickness, i.e. 30 and 10 nm.[31] However, with these previous samples, it was not possible to observe the AP alignment except for the case for s=80 nm lateral separation. This precludes the possibility to systematically investigate the dynamic dipolar coupling between the two layers, when they are magnetized in the opposite direction.

In the present work, we performed a systematic investigation of the SW band structure in Py(10nm)/Cu(5nm)/Py(30nm) trilayer NW arrays with the NW lateral separation ranging from 80 to 280 nm. We found that for this modified Cu thickness, the AP magnetization orientation is stabilized within a specific magnetic field range for all the NW lateral separations thus enabling us to compare the MBS in the P and AP configurations. In addition, a complete understanding of the SW character,

i.e dispersive and non-dispersive modes, has been obtained by calculating the spatial profiles of the modes and the phase relationship between dynamic magnetizations in the two layers.

This has allowed us to highlight the interplay between the **intra**- and **inter**- dynamic dipolar interactions acting within any NW and between NWs, respectively. The former determines, due to the phase relation between the precessing magnetization vectors in the two layers (in-phase or out-of-phase precession), the precession modes which have the largest average dynamic magnetization within one and the same NW. The latter is responsible for the coupling between NWs and the properties of the Bloch collective SWs propagating through the array. It is shown that only the modes characterized by an in-phase precession in the two layers can propagate through the array with appreciable group velocity.

We would like to remark that one specific advantage of the proposed layered structure over planar systems, is that by using a single fabrication mask one can fabricate multilayered structure with different thicknesses of the layers and thus multiple reprogrammable magnetization configurations and dynamic responses. In this respect, layered structures are also interesting as prototypes of multiple magnetic storage layers [32] and as non-volatile vertical magnetic logic gates [33] where the knowledge of the SW properties is important because they represent a source of noise in reading heads [34] as well as a source of energy consumption during the magnetization reversal process.[35] In addition, exploring the third dimension in magnonics is important in order to follow the latest trend in CMOS electronics where expansion from two-dimensional planar to three-dimensional vertically integrated structures is now pursued.[36,37]

## 2. SAMPLE FABRICATION AND EXPERIMENTS

The nanowire arrays were fabricated on a 100 × 100 $\mu m^2$ area using high-resolution electron beam lithography (EBL), electron beam (e-beam) evaporation, and lift-off processes. A 200 nm-thick polymethyl methacrylate (PMMA) EBL resist was first spin-coated on an oxidized Si (001) wafer substrate. The substrate was baked at 180 °C for 90 s on a hot plate. After the exposure, the substrate was developed in a 1:3 solution of methyl isobutyle ketone and isopropyl alcohol (IPA). The multilayer film Py(10nm)/Cu(5nm)/Py(30nm) was then deposited at a rate of 0.2 Å/s on the patterned resist in a deposition chamber with the base pressure of $2 \times 10^{-8}$ Torr or less. This was followed by the lift-off process in acetone. The NW arrays were patterned using a 75 kV lithography system (Elionix ELS 7700) with a resist dose of 512 $\mu C/cm^2$. The NW width is fixed at $w$=280 nm while the NW lateral separation ($s$), and hence the periodicity ($a$), are varied. These parameters and the corresponding widths of the Brillouin zone ($\pi/a$) are reported in Table I. The fabricated trilayer nanowires were examined under a JEOL JSM-6700F field emission scanning electron microscope (SEM). A continuous (unpatterned) Py(10nm)/Cu(5nm)/Py(30nm) trilayer was also prepared and used as a reference sample.

Table I: The geometric parameters of the four nanowire arrays.

| Separation, $s$ (nm) | Periodicity, $a$=$w$+$s$ (nm) | Brillouin zone boundary, $\pi/a$ ($10^7$ rad/m) |
|---|---|---|
| $s_1$=80 | $a_1$=360 | $\pi/a_1$=0.87 |
| $s_2$=100 | $a_2$=380 | $\pi/a_2$=0.83 |
| $s_3$=150 | $a_3$=430 | $\pi/a_3$ =0.73 |
| $s_4$=280 | $a_4$=560 | $\pi/a_4$ =0.56 |

The hysteresis loops were measured in the longitudinal configuration using magneto-optical Kerr effect (MOKE) magnetometry with a photoelastic modulator operating at 50 kHz and lock-in amplification. An in-plane magnetic field $H$ is swept along the NW length. The laser beam has a circular spot with a typical diameter of 100 microns so that the obtained signal was averaged over a large number of NW periods. The incident angle of light on the sample was about 30 degrees.

The SW band dispersions were measured by Brillouin light scattering (BLS) spectroscopy. A beam of monochromatic p-polarized light of wavelength $\lambda$=532nm is focused on the sample surface over an area of about 40 micron in diameter. The sample is mounted on a goniometer to allow rotation around the field direction, i.e. to vary the incidence angle of light ($\theta$). For all the investigated NW arrays, we measured the BLS spectra as a function of the wave vector $k=(4\pi/\lambda)\times\sin(\theta)$, directed perpendicular to the NW length, in the range between 0 and $2.0\times10^7$ rad/m.[38] From the frequency position of the BLS peaks detected at different $k$-values, we were able to map the SW dispersion and to reconstruct the band structure up to the edge of the first Brillouin zone ($\pi/a$), whose values are listed in Table I. A dc magnetic field ($H$), with variable magnitude in the range between -50 and +50 mT, was applied along the sample plane, parallel to the NWs length ($y$-direction), and perpendicular to the scattering plane ($x$-$z$ plane), in the so-called Damon-Eshbach geometry.

## 3. THEORY

In our previous work with a similar sample geometry, we used a Hamiltonian-based method which reproduced both quantitatively and qualitatively the MBS.[31] This model, which is described in detail in Ref. 39, allowed us to obtain the probability amplitude in terms of the spatial dependence of the square root of the total mean-square amplitude for the fluctuating magnetization across the NW width in both the P and AP magnetization configurations. This probability amplitude, however, does not provide information about the relative phases for the modes oscillating in the two ferromagnetic layers, but it does allow us to illustrate important differences in the spatial structure of the modes.

To overcome this limitation of the previous theoretical approach, we also use here an alternative macroscopic (classical) description that conveniently captures the phase difference between the precessing magnetization vectors in the two layers explicitly. In order to calculate the SW dispersion, we employ the numerical code based on the solution of the linearized Landau-Lifshitz-Gilbert (LLG) equation in the magnetostatic approximation.[40,41] We employ a description of the dynamic dipole field of the precessing magnetization in the form of a two-dimensional Green's function which accounts for both the self-demagnetising (dipolar) field of each individual stripe and the collective demagnetising field of the dense array of stripes.[27] Since the sizes of the stripe cross-section are comparable to the exchange length for Permalloy (which is ~5nm), we also include contribution from the effective exchange field of dynamic magnetization. We assume "unpinned surface spins" boundary conditions at all NW surfaces. [42] This is largely because the strong shape anisotropy of the stripe geometry results in the largest contribution to the NW frequency. This is also because of the small layer thicknesses used here, where any perpendicular surface anisotropy, spontaneously induced at stripe upper and lower surfaces during sample fabrication, renormalizes the layer saturation magnetization [43] rather than producing a magnetization surface pinning effect. [44]

The integral operator involving the Green's function and the differential operator of the effective exchange field are discretized on a square mesh, which fills the stripe cross-section. In addition, the interface boundary conditions for the dynamic magnetization are included in the model. In our case, because of the presence of the thick Cu spacer, no interlayer exchange coupling has been considered.[45] The discretization of the operators transforms the LLG equation into an eigenvalue/eigenvector problem for a matrix. The eigenvalues of the matrix represent eigen-frequencies of the Bloch SW modes for the periodic stripe array for a given Bloch wave number, and the eigen-vectors are the spatial profiles for the respective modes. The eigen-vectors are complex-valued; thus, they contain information about the phase of magnetization oscillation at any particular point on the nanostructure cross-section.

The constructed eigenvalue/eigenvector problem is solved numerically using the numerical tools built into MathCAD. A matrix is built for a given value of the Bloch vector then its eigen-values are numerically found. Repeating this procedure for a range of Bloch wave vectors produces a dispersion graph. Calculating eigen-vectors of the matrix in the same program run allows us to visualize the respective SW mode profiles, including the oscillation phase. The mode frequencies delivered by these calculations are very close to those obtained with the microscopic Hamiltonian-based theory. [22] More generally, the consistency between the microscopic and macroscopic approaches to SW dynamics in nanostructure arrays has been discussed in Refs. 39, 46.

## 4. RESULTS AND DISCUSSION

### a. Continuous trilayered films

As a first step in our dynamical characterization, we have measured the MOKE loop and the field and wave vector dependence of the SW frequency for the continuous (unpatterned) Py(10nm)/Cu(5nm)/Py(30nm) film. For this trilayered film, the measured MOKE loop, which is shown in Fig. 1(a), has a square shape with a coercive field of a few tenths of one mT and with almost 100% remanence. There was no evidence of separate switching of the magnetizations in the two layers, which is expected since MOKE mainly probes the magnetization of the topmost (30 nm thick) portion of the film.

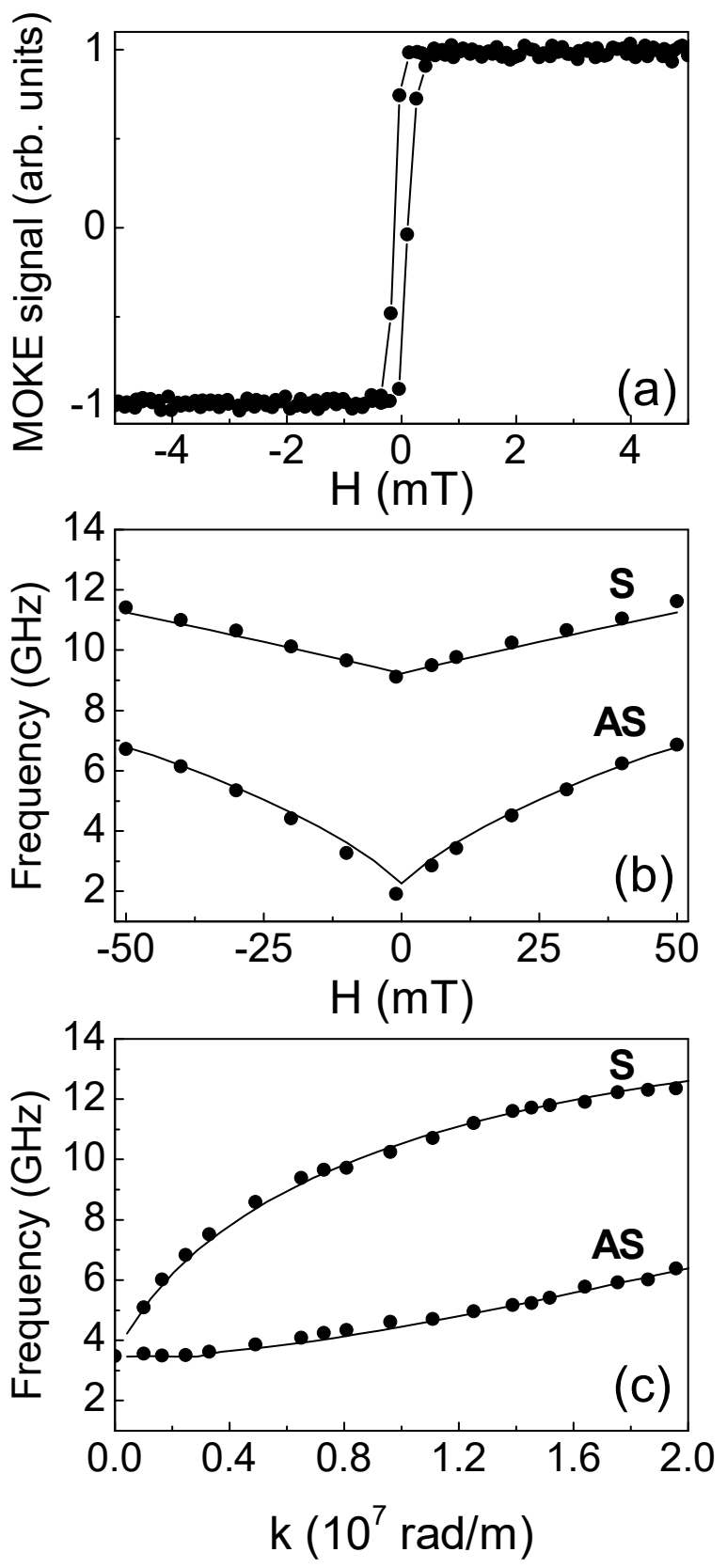

Fig. 1 (a) Longitudinal MOKE loop, (b) magnetic field (with *k* fixed at $0.81\times10^7$ rad/m) and (c) wave vector dependence (with *H* fixed at +15 mT) of the SW frequency for the continuous Py/Cu/Py trilayer. The two observed modes are labeled as symmetric (S) and anti-symmetric (AS) depending on the phase relation (in-phase and out-of-phase) between the dynamic magnetizations in the two Py films.

Fig. 1(b) presents the BLS frequency evolution as a function of the applied magnetic field for a fixed wave vector $k$=0.81×10$^7$ rad/m while Fig. 1(c) shows the dispersion (frequency vs wave vector) for an applied field $H$= +15mT. In both cases, we detected a couple of peaks which correspond to the symmetrical (acoustic) or anti-symmetrical (optic) modes, associated with the in-phase and out-of-phase precession of the dynamic magnetizations in the two Py layers.[47] For this reason, the two modes are labelled as S and AS, respectively. In Panel (b), the mode frequencies exhibit the typical V-shape monotonic evolution vs $H$ while in Panel (c) the S mode has a larger group velocity than the AS mode in agreement with the results of a previous investigation on a layered structure with same thicknesses. [48]

A good agreement between the calculated and measured frequency has been obtained using the following magnetic parameters: exchange stiffness $A = 1.0\times10^{-11}$ J/m, saturation magnetization $M_s$ = $0.73\times10^6$ A/m and $\gamma/2\pi$ = 29.3 GHz/T (with $\gamma$ being the gyromagnetic ratio) These parameters were deduced from a fit to the dispersion of the SW frequency for the continuous Py/Cu/Py trilayered films whose results are presented as solid lines in Fig. 1 (b) and (c).

b. Py/Cu/Py nanowire arrays

In Fig. 2 we present the SEM images (left column) and the MOKE loops (right column) of the NW arrays with different lateral separations ($s$). SEM images reveal well-defined NW arrays with sharp edges and uniform separation. Regarding the MOKE loops, we start by discussing the M-H loop for the $s$=280 nm NW array. Increasing the external field from the negative values, with all the NWs saturated along the field direction, results in a sharp increase in magnetization at around zero field which we attribute to the reversal of the 30 nm thick Py layer. This increase in magnetization is proportional to the volume (or layer thickness) fraction of the NWwhich has reversed its magnetization, given by 30/(30+10).[49] This corresponds to an increase of 75%, which is in good agreement with the values obtained from the $M$-$H$ loops and thus corroborating our assumption that the thicker Py layer reverses first. Further increase in the applied field results in a second and much smaller increase in magnetization at $H$=25 mT. This behaviour reflects on the dM/dH curve which displays two distinct peaks, corresponding to the individual switching of the two layers of different thicknesses within the same NW. In the field range between 0 and 25 mT, the magnetization vectors of the Py layers within the same NW point in opposite directions thus realizing an AP configuration. When the value of $s$ is reduced, we notice significant changes in the measured loop shape. The first jump of the magnetization magnitude (located close to the zero field) tends to be less pronounced, while the second one becomes more intense. In addition, the field range for the AP configuration reduces significantly (to 18 mT) for $s_1$=80 nm. While this latter effect is connected to the different reversal field induced by the magnetostatic effects, the former one is related to the fact that in MOKE experiments, light probes only a thin layer (≈20 nm). Therefore, it provides information about the reversal of the topmost Py layer and the magnetization curve is a convolution of the loop with a decaying optical field. In addition to this, on reducing *s*, a shadowing effect starts to play an important role when light impinges at an oblique angle with respect to the sample normal.

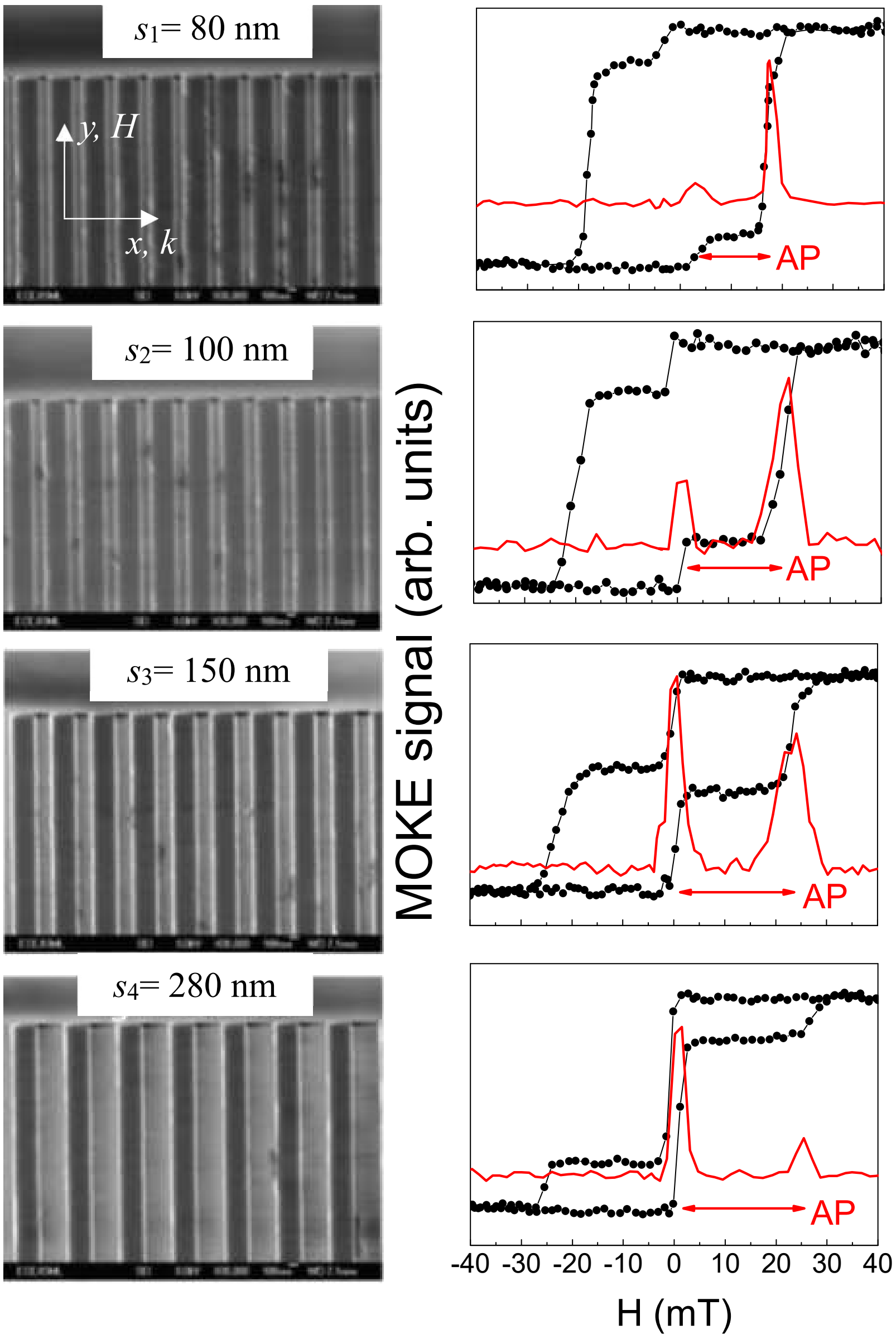

Fig. 2 (Left column) SEM images and (Right column) longitudinal MOKE loops for the Py(10nm)/Cu(5nm)/Py(30nm) trilayer NW arrays with different lateral separation (*s*). The red curves represent the first derivative of the ascending branch of the loop with respect to *H*. Red arrows indicate the field range of the AP alignment of the two layer magnetizations while in the remaining field range the magnetizations are in the P configuration.

In order to identify the field range where the two layer magnetizations are in opposite directions, and to eliminate the effect of the finite penetration depth of light in the MOKE experiment, we have measured the field dependence of BLS spectra for a fixed $k$=0.81×10$^7$ rad/m. Here we emphasize that in the BLS measurements, even if the inelastically scattered light mainly originates from the topmost Py layer, the measured frequencies of the BLS peaks are determined by the entire layered stack (the thickness of all the layers), and these SW frequencies reflect the relative magnetization orientation within the two layers.[47] Therefore, BLS provides, through the frequency behaviour of the detected modes, complementary and more precise information on the magnetic field range where the AP configuration occurs.

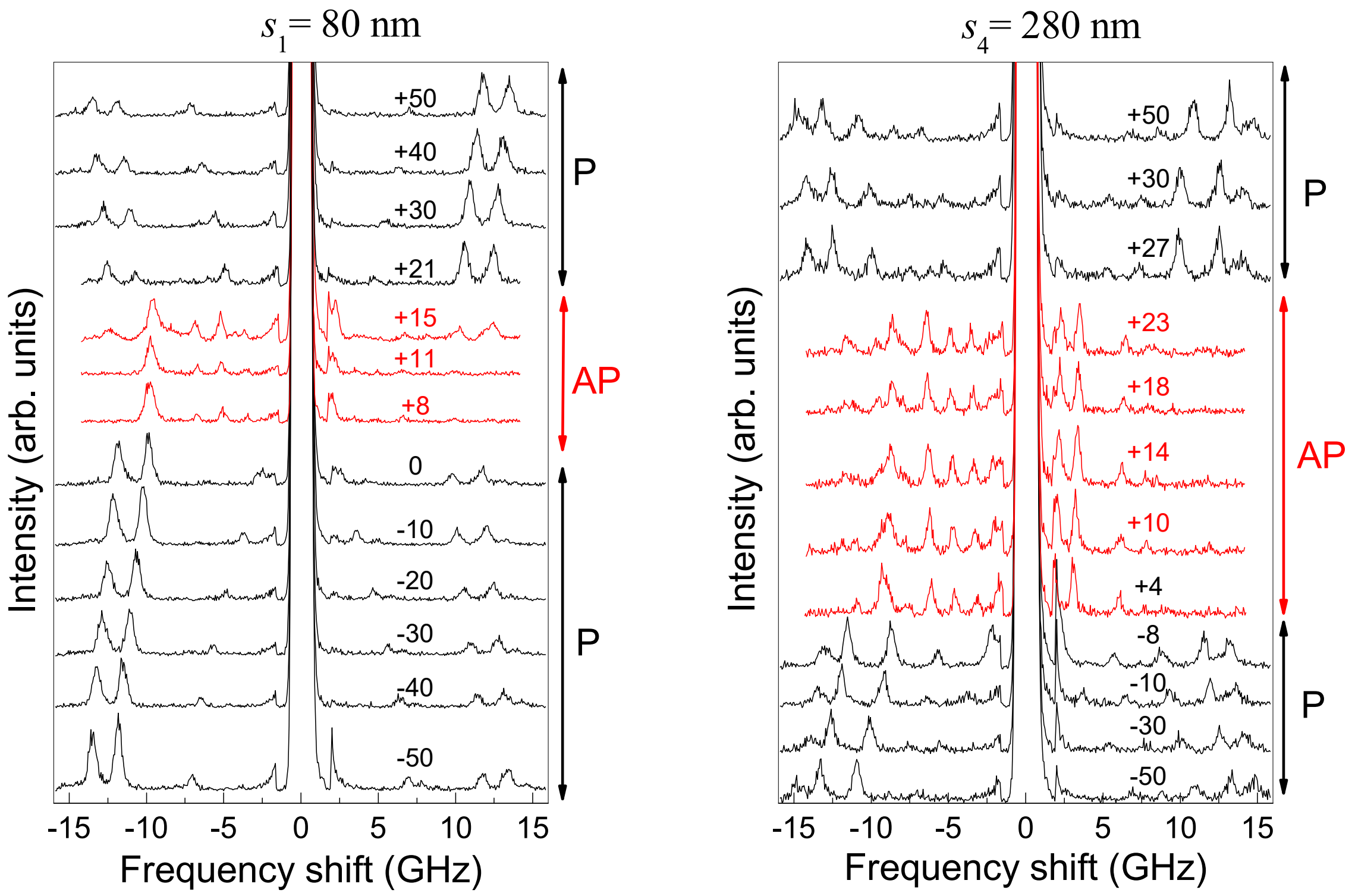

Fig. 3 A sequence of measured BLS spectra for $s_1$=80 (left panel) and $s_4$=280 nm (right panel) as a function of the applied magnetic field $H$ expressed in mT, which is swept from negative to positive values following the ascending branch of the hysteresis loop. Thus it encompasses both the P (black spectra) and AP (red spectra) magnetization configurations.

Fig. 3 shows a sequence of measured BLS spectra for H ranging from -50 to +50 mT for two of the NW arrays, specifically those with the smallest ($s_1$) and the largest ($s_4$) lateral separation. As the field is swept from the negative saturation to zero, with the NW always in the P state, the peaks positions monotonically decrease (black spectra). When the field is reversed, an abrupt variation in the general shape of the red spectra, corresponding to the transition to the AP state, is observed. Then the peak positions remain almost unchanged until the positive field reaches a magnitude needed to co-align the magnetization vectors in the two layers. From this point onwards, the peaks move again to a higher frequency. We notice that the field range for the AP state for the $s_4$ array is larger than that for $s_1$ array.

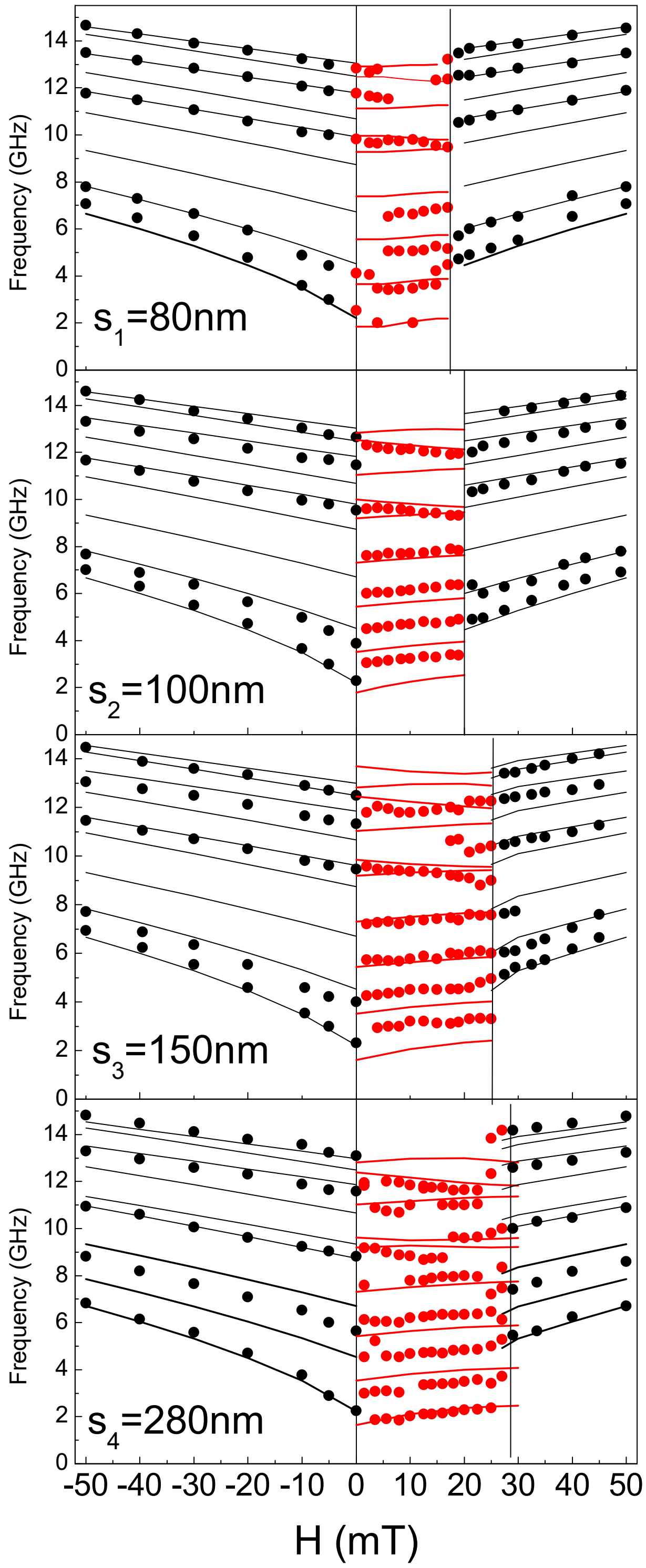

Fig. 4 Field dependence of the measured frequencies, for all the investigated arrays with different separation, in the range of $H$ from -50 to +50 mT. H is swept from negative to positive values thus encompassing both the P and AP magnetization configurations of the Py layers. Black (red) points indicate the frequency values in the P (AP) state. The incidence wave vector $k$ is fixed at $0.81 \times 10^7$ rad/m.

Fig. 4 displays the corresponding frequencies plotted as a function of the applied field magnitude. Starting from $H = -50$ mT, the field is decreased in magnitude down to zero and then reversed to positive values and increased. Therefore, it follows the ascending branch of the hysteresis loops (see Fig. 1). Up to five peaks are experimentally detected in the P state while their number increases between six and eight for the AP configuration. The modes detected in the P configuration monotonically evolve as a function of the applied magnetic field. At the P to AP transition, there is an abrupt frequency jump, and the frequencies remain roughly constant (or slightly increase for the lower modes) with $H$ in the entire AP range. Once the positive saturating field has been reached, the frequencies follow again an almost linear dependence on $H$. An overall very good agreement between the calculated and measured frequencies is obtained in the whole field range investigated. From the BLS measurements, the field range where the AP state is observed is slightly larger than that seen as the almost flat plateaux in the MOKE loops.

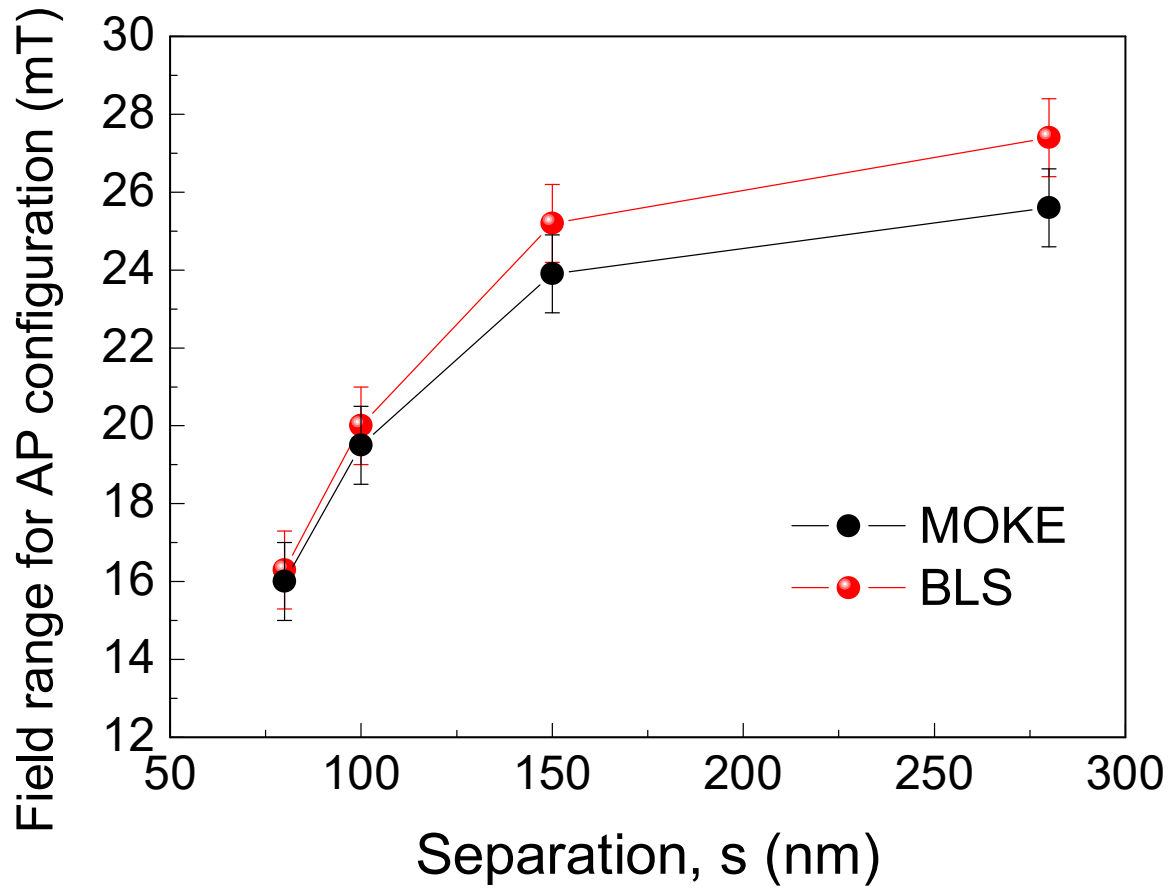

Fig. 5 Field range versus separation (s) for the AP configuration as derived by MOKE (black points) and BLS (red points) measurements. The lines are guides to the eye.

The results for the field range versus separation of the AP configuration as obtained by BLS and MOKE measurements are summerized in Fig. 5. It is evident that a significant increase is observed up to 150 nm of lateral separation followed by a gradual approach to saturation for larger separation values.

Fig. 6 displays the comparison between the measured and the calculated SW dispersion within the first BZ for NW arrays with different separations (*s*) and in the P-state (left column) and AP-state (right column) magnetization configurations. When plotted over the first BZ, the bandwidth of the most dispersive mode shows the typical periodic Bloch-type behaviour induced by the artifical periodicity of the lattice.[50]

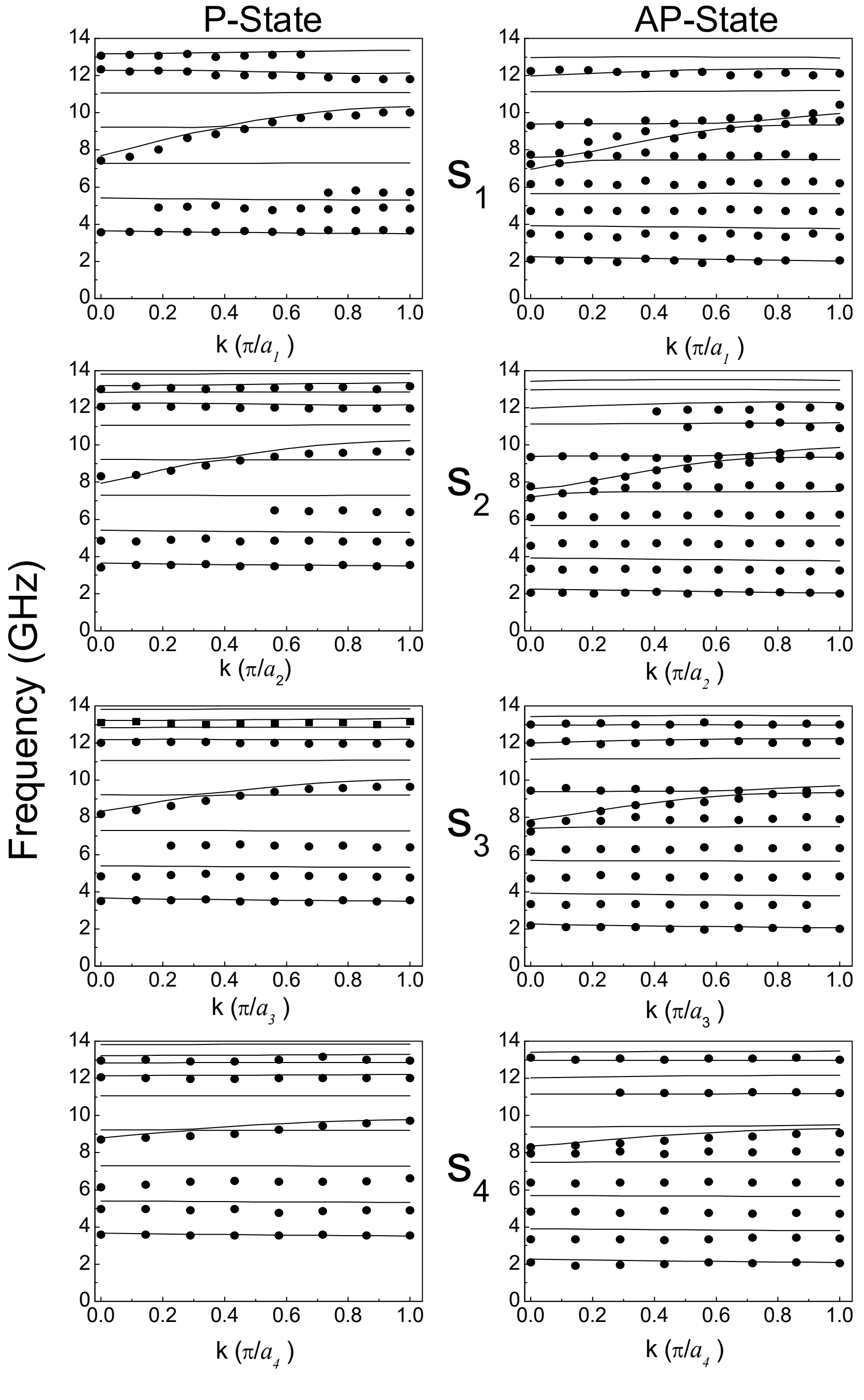

Fig. 6 Comparison between the measured and calculated band structure in the first Brillouin zone for NW arrays with different separations (s) at *H*= + 15mT in the P (left column) and AP (right column) magnetization configuration.

While most of the modes are dispersionless, i.e. their frequencies do not change as a function of the wavevector, there is always one mode which exhibits a significant magnonic bandwidth, whatever the magnetization configuration (P or AP) may be. By contrast with the case of single-layer nanowire arrays, where only the lowest frequency mode exhibits sizeable dispersion, in the present case of layered NWs, the dispersive mode is at intermediate frequencies within the MBS while there are two (or sometimes more) non-dispersive modes at lower frequencies. As a general comment, on increasing the separation, a reduction of the dispersive mode group velocity is seen together with an overall MBS downshift of a couple of GHz. A larger number of modes is observed in the AP state rather than in the P state. We also see the presence of a mode hybridization occurring at different *k* and frequency values where the dispersive mode interacts with a stationary mode giving rise to a mode repulsion

To summarize the principal magnonic effects derived from Fig. 4, we give in Fig. 6 a comparison between the measured (points) and calculated (lines) magnonic bandwidths for the most dispersive modes as a function of the NW separation (*s*) for both the P and AP configurations. In these cases, the bandwidth decreases on increasing *s* and, most importantly, the value measured in the AP state is always smaller than that for the P configuration. This suggests a reduction of the interlayer dipolar coupling when the layers magnetization vectors point in opposite directions. The reduction is due to magnetization vectors in the two layers precessing in opposite senses. In this case, only one vector component of dynamic magnetetization can be in-phase in both layers at any time. The other component is then in anti-phase, and hence does not produce contribution to a long-ranging (global) dynamic dipole field coupling neighbouring NW on the array.

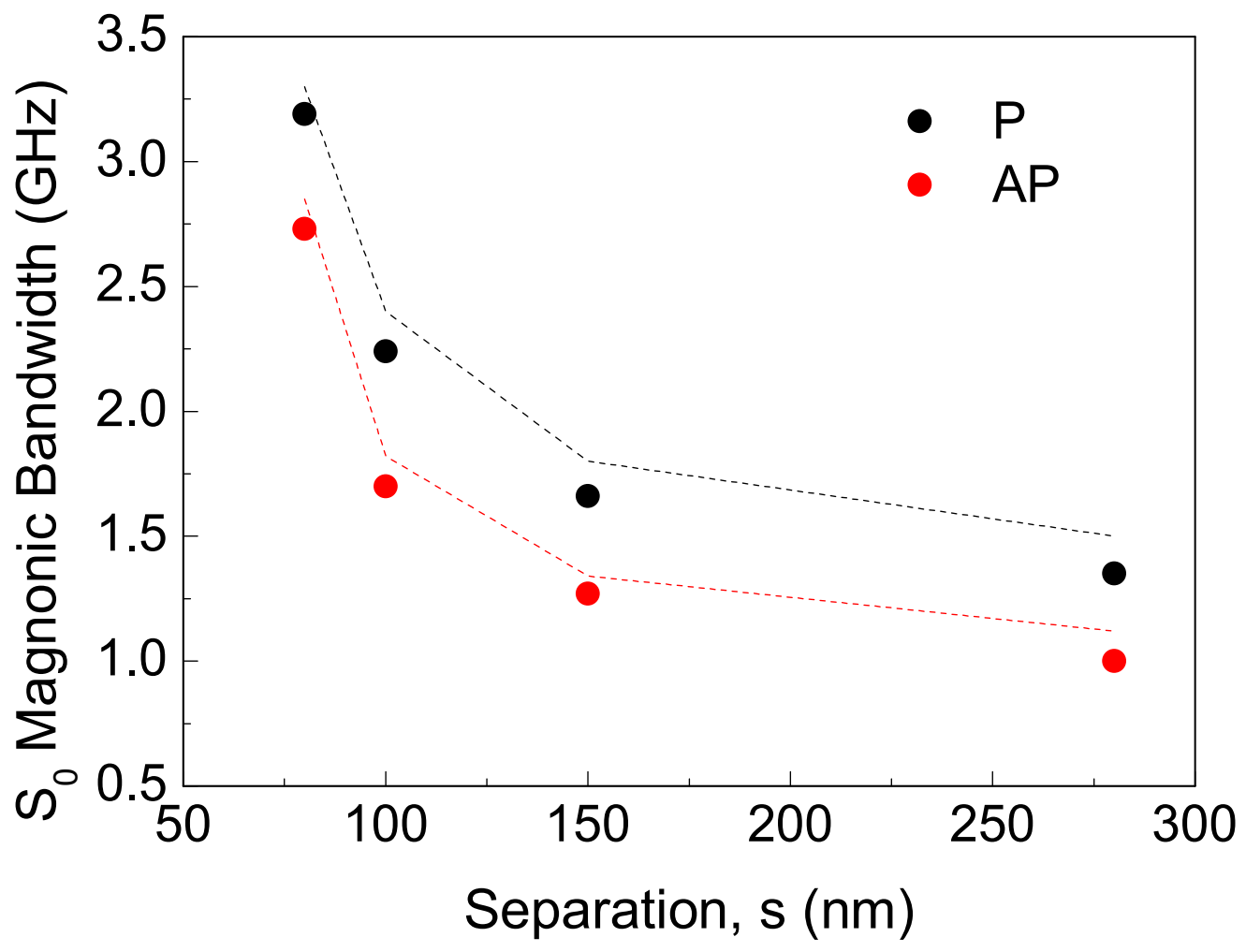

Fig. 7 Comparison between the measured (points) and the calculated (curves) magnonic bandwidth of the dispersive mode as a function of the four NW separations in the P and AP states. The straight sections of lines are guides to the eye.

To explain the properties of the measured SW band structure and in particular to understand the difference in the $k$-dependences for the modes, i.e. whether stationary or dispersive in character, we study the spatial distribution of the dynamic magnetization across the NW width calculated for $k$=0 and $k=\pi/a$. These simulations are carried out for the P and AP states and for an external field of $H$= +15 mT.

In Figs. 8 and 9, we present the calculated amplitude distributions for the eight lowest frequency modes in the two Py layers in the P and AP states, respectively. These data were collected for the NW array with separation $s_1$. Similar results have been obtained for the other samples. In these graphs, the profiles are averaged over the thickness of the respective layer. This is a valid procedure, since the dynamic magnetization is almost uniform across the thickness of each layer for all displayed profiles. Red and black curves represent the in-plane component of the dynamic magnetization in the NW layers while the blue curve is the averaged value of the individual ones.

In all cases, the mode profiles are very close to the sinusoidal shape and are characterized by an increasing number of nodes across the NW width. Hence, one deals with standing SWs bouncing between the edges of each layer.

Furthermore, the two ferromagnetic layers of each NW can be regarded as a system of two resonators which are coupled by the dipole interactions (interlayer exchange interaction is here neglected, because the inter-layer spacer thickness is too large). Therefore, similarly to what made for the continuous trilayer, the family of the resonance modes can be separated into acoustic (S) and optical (AS) modes depending on whether the oscillation in the two layers is in-phase or in anti-phase.

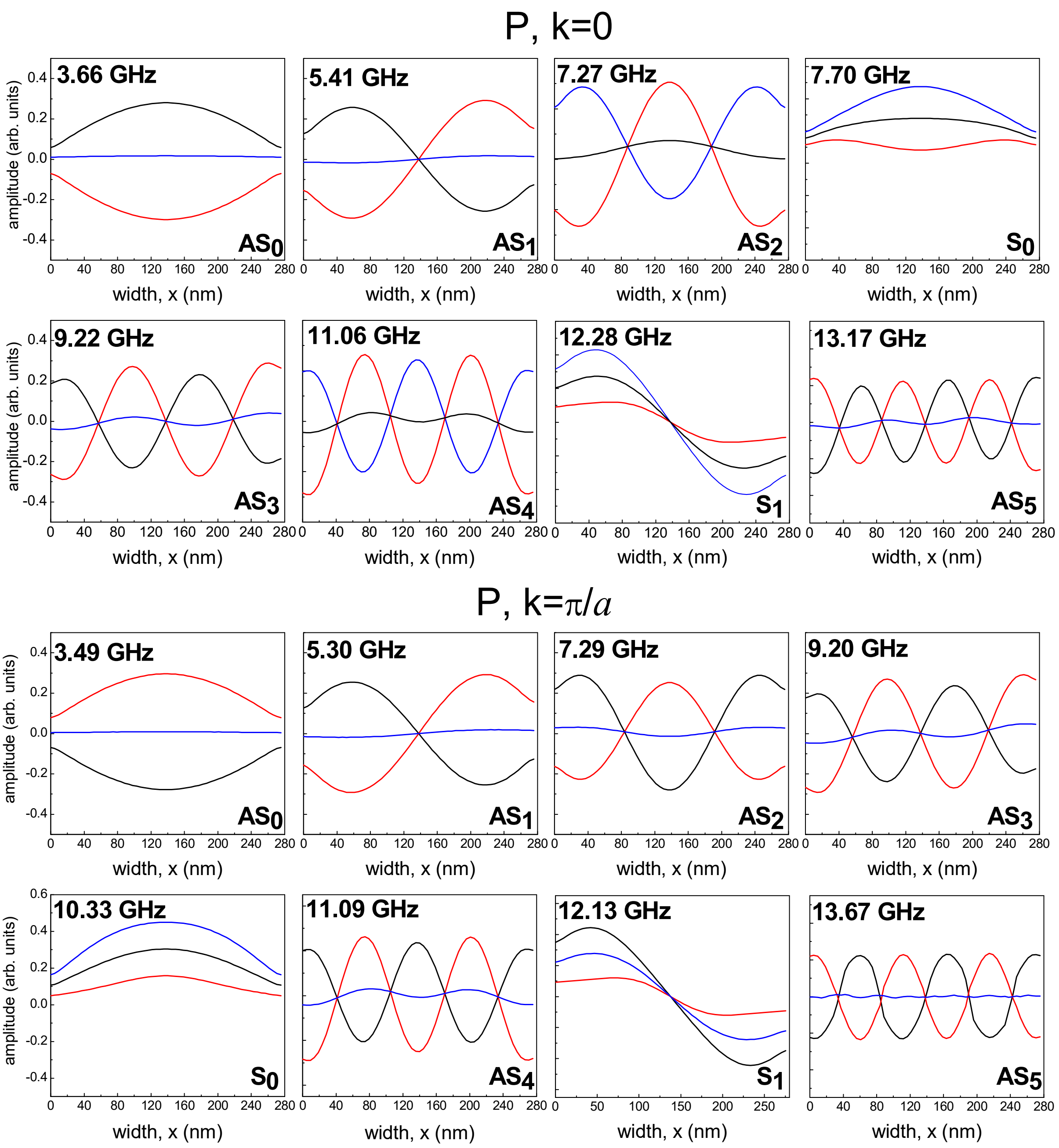

Fig. 8 Spatial distribution (in-plane *x*-component) of the eight lowest-frequency modes for the Py/Cu/Py NWs for $s_1$=80 nm calculated along the NW width (*x*-direction) and averaged over the thickness (z-direction) for the P state at $k$=0 (upper panel) and $k=\pi/a$ (lower panel). The magnetic field $H$= +15 mT is applied along the NW length. Black (red) curves refer to the thick (thin) Permalloy layer while the blue curves are the average value of the dynamic magnetization across the two layers. Modes are labelled as S (in-phase) and AS (out-of-phase) depending on the relative phase of magnetization precession in the two layers. The subscript number indicates the number of nodes for each mode.

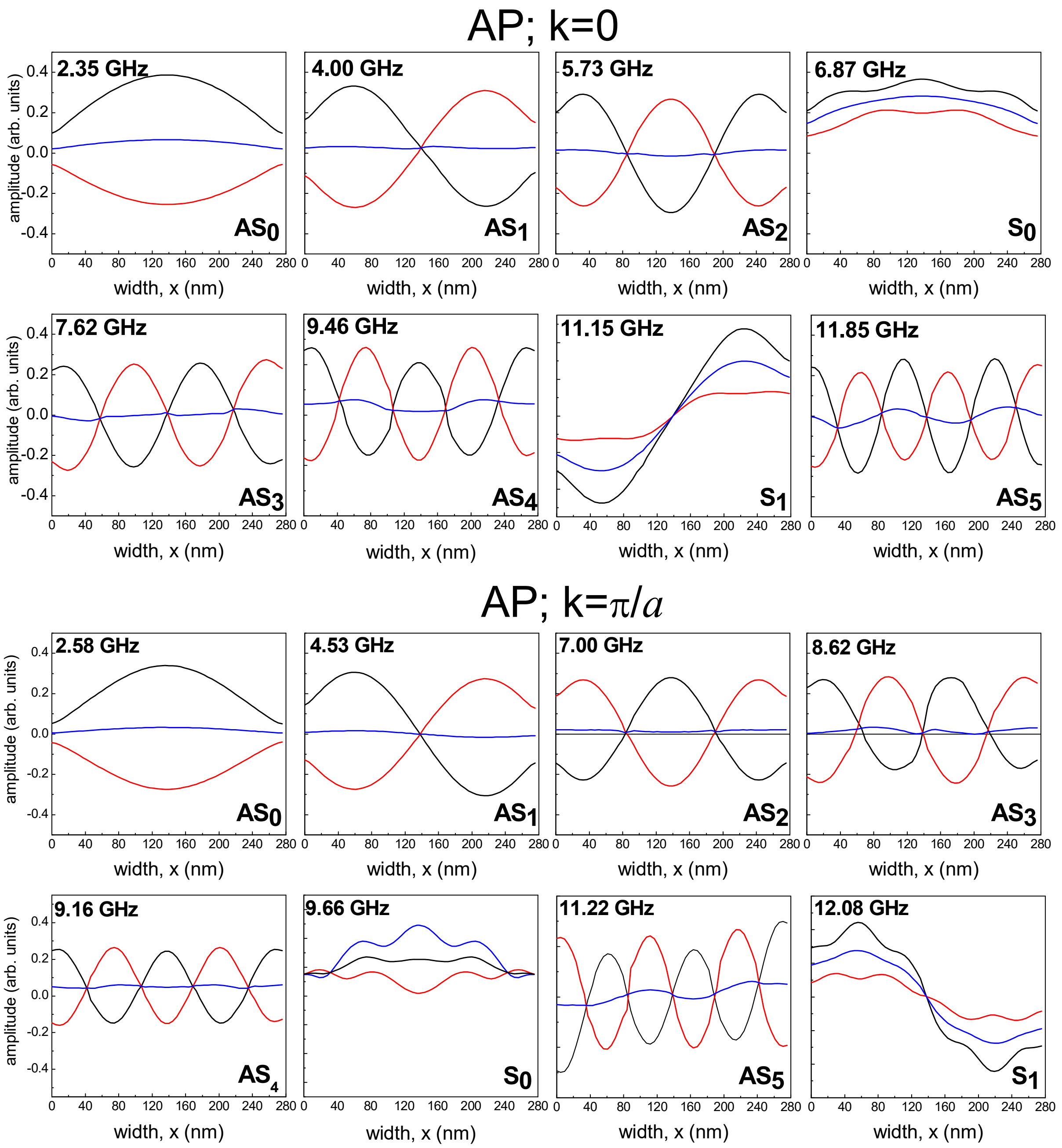

Fig. 9 The same as in Fig. 8, but for the AP state.

Let us first discuss the modes in the P and AP configurations at $k$=0. In this case, the two lowest frequency modes belong to the AS family of modes, having either zero or one node within the NW width, respectively. The former ($AS_0$), being characterized by a nearly uniform precession amplitude across the width of the NWs, can be considered as the quasi-uniform (fundamental) mode of the NWs while the $AS_1$ has one node in the middle of the NW width for each layer. They are non-dispersive. This is because the magnetization vectors precession is in anti-phase in the two layers and, consequently, the mean dynamic magnetization amplitude within the same NW is close to zero (relative to the individual maximum amplitudes). For this reason, although the intensity of their BLS peaks is not negligible, because light essentially probes the top layer, only, their dynamical coupling through the array is not strong enough in order to permit the formation of collective traveling modes of Bloch type on the periodic array. This is a peculiar feature of the layered NWs since in the case of single layer NW this lower frequency mode is the one showing the largest magnonic bandwidth.

A different situation occurs for the S modes where the average magnetization is significantly different from zero (i.e. comparable with the maximum amplitudes in the individual layers). The magnetization oscillations in the two layers are now in-phase. This creates a strong and far reaching dynamic stray field outside of the NW itself. As a result, the neighbouring NWs are efficiently dipole coupled. This gives rise to collective SWs of the Bloch type. This mainly happens for mode $S_0$, which is the quasi-uniform mode of the NW layers. The coupling decreases on increasing the number of nodes, as can be seen for modes $S_1$ and $S_2$. Interestingly, the frequency of the $S_0$ mode is the third smallest for the P state, while it is the fourth smallest in the AP configuration. Similarly, modes $S_1$ and $S_2$ are located differently in frequency with respect to the other AS modes for the two states. Regarding the mode profiles calculated at $k=\pi/a$, they are not significantly different from those at $k=0$ in always having sinusoidal profiles across the NW width. The most significant difference is the frequency position of the $S_0$ mode in the P state. Due to its dispersive character (its frequency increases by more than 2 GHz on passing from $k=0$ to $\pi/a$), it becomes the fourth lowest frequency mode at $k= \pi/a$ while it was the third in the $k=0$ state.

Similar considerations hold for the AP state (see Fig. 9), where the most dispersive mode - the $S_0$ changes its frequency position with respect to the non-dispersive AS modes. At $k=0$ the $S_0$ mode is the fourth from the bottom of the spectrum for both the P and AP state while, at $k=\pi/a$, it is the fifth for the P state and the sixth in the AP state. In addition, it is worth mentioning a slight deviation of the mode profiles from being a pure sine wave; there is a modulation of the sine wave with a harmonic function of a different period that is observable in the profiles. These results enabled us to interpret the measured dispersions plotted in Fig. 6 for all the investigated NW arrays.

## CONCLUSIONS

We have investigated both experimentally and theoretically the dependence of the spin wave band structure in Py/Cu/Py nanowire arrays for different values of the separation in the range between 80 and 280 nm. In contrast to a single nanowire array, the layered structure offers the possibility to stabilize the anti-parallel configuration for the magnetization vectors in the two Py layers over different magnetic field ranges whose values depend on the nanowire separation.

Significant differences in the magnonic band structure, including the bandwidth and the number of dispersive and stationary modes, have been observed when comparing the parallel (P) and anti-parallel (AP) configurations. The obtained results have been successfully explained quantitatively and qualitatively by calculating the frequency dispersion of the different modes as well as their spatial profiles. In this regard, a thorough understanding of the individual mode character (stationary or dispersive) is explained by considering the phase relation (in-phase or out-of-phase) between the magnetization oscillations in the two Py layers within each nanowire.

We believe that our work is a significant step forward in understanding the spin wave band structure and its field reprogrammability and will stimulate very active research on layered and 3D magnonics in the near future.